\documentclass[aps,pra,pdf,superscriptaddress,twocolumn,showpacs,nofootinbib]{revtex4-2}
\usepackage{amsmath,amssymb,amsfonts}
\usepackage{eqnarray}
\usepackage{graphics,float}
\usepackage{bm}
\usepackage{braket}
\usepackage{amsmath}
\usepackage{physics}
\usepackage{mathtools}
\usepackage{gensymb}
\usepackage{color,xcolor,colortbl}
\usepackage{tikz}
\usepackage{soul}
\usetikzlibrary{shapes}
\usepackage[colorlinks=true,
linkcolor=blue,
filecolor=magenta,      
urlcolor=blue,
citecolor=blue]{hyperref}

\definecolor{Gray}{gray}{0.85}
\definecolor{LightCyan}{rgb}{0.25,0.9,0.95}
\definecolor{LightMagenta}{rgb}{0.91,0.21,0.96}

\graphicspath{{Plots/}}

\begin{document}

    \title{Dynamical signatures of superfluidity and shear rigidity in different phases of a dipolar Bose-Einstein condensate}
	
	\author{Soumyadeep Halder}
	\email{soumya.hhs@gmail.com}
	\affiliation{Department of Physics, Indian Institute of Technology Gandhinagar, Palaj, Gujarat 382055, India}
	
	\author{Hari Sadhan Ghosh}
	\affiliation{Department of Physics, Indian Institute of Technology Kharagpur, Kharagpur 721302, India}
	
	\author{Axel Pelster}
	\affiliation{Department of Physics and Research Center OPTIMAS, Rheinland-Pf\"{a}lzische Technische Universit\"{a}t Kaiserslautern-Landau, Kaiserslautern, Germany}
	
	\author{B. Prasanna Venkatesh}
	\email{prasanna.b@iitgn.ac.in}
	\affiliation{Department of Physics, Indian Institute of Technology Gandhinagar, Palaj, Gujarat 382055, India}
	
	\begin{abstract}
		We show that a sudden change in the polarization direction of the magnetic dipole moments of the atoms in a dipolar Bose-Einstein condensate (BEC) can serve as a useful dynamical probe to sense its superfluid and solid-like properties. We find that for small angular deviation of the polarization direction, actuated for instance by modifying an external magnetic field, the superfluid state undergoes an undamped scissors mode oscillation, a characteristic signature of superfluidity. In contrast, both the droplet and supersolid states exhibit a scissors-mode oscillation, which is effectively damped due to multiple closely spaced frequency components. Notably, we find that this damping rate provides a qualitative measure for the rigidity of different phases of a dipolar BEC. Furthermore, there exists a maximum angular deviation of the polarization direction, beyond which the droplet and the supersolid states undergo a permanent deformation i.e., we find an analog of the usual elastic to plastic phase transition of solids. We characterize this transition numerically using the fidelity of the condensate wavefunction with the ground state as well as the droplet width and periodicity of the supersolid density of the condensate which are experimentally accessible. Thus, the dynamical protocol introduced here can be an important experimental benchmark to identify and characterize the superfluid and solid properties of different phases of dipolar BECs.
	\end{abstract} 
	
	\maketitle
	
	\section{Introduction}
	In recent years, dipolar Bose-Einstein condensates (BEC) have attracted significant attention and broadened the scope of potential research directions in the field of ultracold quantum gases, featuring both long-range anisotropic dipolar and short-range isotropic contact interactions. In addition to the characteristics of non-dipolar BECs, a dipolar BEC exhibits a range of intriguing quantum phenomena, such as anisotropic superfluidity \cite{ticknor_Anisotropic_2011, wenzel_Anisotropic_2018, seo_Mirror_2024}, the quantum analog of Rosensweig instability \cite{kadau_Observing_2016}, the emergence of roton excitations \cite{santos_RotonMaxon_2003, fischer_stability_2006, chomaz_Observation_2018, petter_Probing_2019, schmidt_Roton_2021}, and the formation of quantized vortices through magneto-stirring \cite{klaus_Observation_2022}. The interplay between the long-range anisotropic dipolar and short-range isotropic contact interaction gives rise to different unique ground state phases \cite{wachtler_Groundstate_2016, bisset_groundstate_2016, cinti_Classical_2017, baillie_Droplet_2018, mishra_SelfBound_2020, zhang_Phases_2021, zhang_Metastable_2024, he_Accessing_2025, sanchez-baena_Tilted_2025}. The long-sought supersolid and self-bound quantum droplet states were also recently observed in dipolar BEC of highly magnetic Dy and Er atoms \cite{ferrier-barbut_Observation_2016, chomaz_QuantumFluctuationDriven_2016, bottcher_Transient_2019, chomaz_LongLived_2019, tanzi_Observation_2019}, where the dominant dipolar attraction-driven collapse is stabilized by the effect of quantum fluctuations \cite{petrov_Quantum_2015}. \par 
	
	Supersolids are an intriguing state of matter formed by the spontaneous breaking of the gauge symmetry, as in a superfluid, and continuous translational symmetry like a solid \cite{leggett_Can_1970,watanabe_Spontaneous_2012, fischer_identifying_2012, leonard_Supersolid_2017}. Following the experimental realization of supersolids in atomic dipolar BECs \cite{bottcher_Transient_2019, chomaz_LongLived_2019, tanzi_Observation_2019, tanzi_Supersolid_2019}, the research on supersolids and self-bound droplets has experienced a surge of interest \cite{norcia_Twodimensional_2021, chomaz_Dipolar_2022, bland_TwoDimensional_2022, ripley_Twodimensional_2023, he_Observation_2025}, extending to diverse systems including binary dipolar BECs \cite{trautmann_Dipolar_2018, bisset_Quantum_2021, smith_Approximate_2021, smith_Quantum_2021, politi_Interspecies_2022, halder_Twodimensional_2023, scheiermann_Catalyzation_2023, halder_Induced_2024, kirkby_Excitations_2024, he_Quantum_2024,  scheiermann_Excitation_2025} and polar molecular BECs \cite{bigagli_Observation_2024, ciardi_SelfBound_2025, langen_Dipolar_2025, langen_Quantum_2024, stevenson_ThreeBody_2024, zhang_Observation_2025, jin_BoseEinstein_2025}, as well as ion-doped helium nanodroplets \cite{matos_Transitional_2025}. On the other hand, the ability to tune the interatomic interaction to characterize the properties of different phases of dipolar BECs has spurred extensive theoretical and experimental research, including measurement of superfluid fraction \cite{tanzi_Evidence_2021, gallemi_Superfluid_2022, chauveau_Superfluid_2023, mukherjee_Supersolid_2023}, global phase coherence \cite{tanzi_Observation_2019, chomaz_LongLived_2019, ilzhofer_Phase_2021}, Josephson oscillation \cite{biagioni_Measurement_2024}, excitation of breathing and scissors mode oscillation \cite{ferrier_barbut_scissors_2018, roccuzzo_Moment_2022, zhen_Breaking_2025}, Berezinskii-Kosterlitz-Thouless (BKT) transition \cite{bombin_BerezinskiiKosterlitzThouless_2019, he_Exploring_2025}, formation of quantized vortices \cite{gallemi_Quantized_2020, klaus_Observation_2022, sindik_Creation_2022, casotti_Observation_2024, halder_Roadmap_2025, das_Unveiling_2025} and persistent current \cite{tengstrand_Persistent_2021, ghosh_Induced_2024, mukherjee_Selective_2025}. However, most of these investigations so far have only demonstrated the superfluid aspects of the supersolid phase. The fundamental properties of solids in a supersolid have not been extensively investigated. The solid nature of the supersolid is inferred only from the crystalline arrangements of the droplets and density peaks at finite momenta of the density distribution in the momentum space \cite{natale_Excitation_2019, petter_Bragg_2021}. Interestingly, these droplets are also made of superfluid and individually showcase phase coherence. This leads to a basic curiosity \cite{josserand_Coexistence_2007, kuklov_How_2011}: To what extent are supersolids and quantum droplets truly solid?\par 
	 
	Elasticity is a fundamental property of solids that describes how the system responds to the externally applied forces and their ability to return to their original shape and size after the external force is removed. Growing attention has been directed toward understanding the elastic properties of dipolar supersolids and isolated droplet states. Most theoretical studies to date \cite{blakie_Compressibility_2023, platt_Sound_2024, poli_Excitations_2024, sindik_Sound_2024} estimate the elastic parameters either from the ground state energy density or by analyzing the excitation spectrum obtained by linearization about the ground state configuration. These methods offer only an indirect static description of elasticity, capturing the equilibrium stiffness of the system under small perturbations. Consequently, they do not account for possible elastic to plastic transitions that may arise under strong external perturbations. In contrast, a dynamical probe directly explores the real-time response of the system to external perturbations, providing a more comprehensive understanding of its elastic behavior. Such approaches not only reveal the collective excitation modes associated with both the crystalline and superfluid degrees of freedom but also establish a direct connection to experimentally measurable quantities. Although a few very recent studies have begun to explore the dynamical elastic response of supersolids, they primarily rely on perturbations of the trapping potential \cite{zhao_Elastic_2025, bougas_Signatures_2025, senarathyapa_Anomalous_2025}. While these approaches have provided important insights, they probe elasticity indirectly, as trap modifications alter the density profile, which subsequently influences the interaction strength in an implicit manner. Since elasticity fundamentally originates from interparticle interactions, this leaves an important gap in the literature: a direct dynamical assessment that systematically explores the role of interparticle interactions in governing elasticity in different phases of dipolar BECs.\par
	
	On the other hand, measurements of the scissors-mode frequency have long been employed as a diagnostic tool for exploring superfluidity across various systems \cite{bohle_New_1984, marago_Observation_2000, wright_FiniteTemperature_2007, bismut_Collective_2010, vanbijnen_Collective_2010, qu_Scissors_2023}. In the conventional scissors mode excitation scheme, involving a sudden rotation of the trap \cite{guery-odelin_Scissors_1999, marago_Observation_2000}, the frequency is determined solely by the trap anisotropy and is independent of the interatomic dipole–dipole interaction (DDI) \cite{bismut_Collective_2010, roccuzzo_Rotating_2020, roccuzzo_Moment_2022}, thus failing to capture the elastic properties of the system. Furthermore, although in quasi-1D geometries the scissors mode frequency has been theoretically shown to characterize the superfluid nature of different phases of dipolar BECs \cite{roccuzzo_Rotating_2020, tanzi_Evidence_2021}, the situation differs in quasi-2D systems with crystalline ordering \cite{young-s._Dipolemode_2023}. Both numerical simulations and experiments demonstrate that, in such cases, the scissors-mode frequency remains nearly unchanged even when the superfluid fraction is significantly reduced i.e., during the transition from a coherent superfluid to an incoherent array of isolated droplets \cite{norcia_Can_2022}. In fact, the frequency remains close to the superfluid prediction and hence deviates far from that expected for isolated droplets. Thus, the conventional method of exciting the scissors mode does not always provide a reliable experimental probe of superfluidity, especially in the case of quasi-2D dipolar BEC phases.\par 
	
	In this article, we demonstrate a protocol to dynamically probe both superfluidity as well as rigidity in different phases of a quasi-2D dipolar BEC by applying a sudden change in polarization direction as schematically depicted in Fig. \ref{fig:schematic}. Within the small angular deviation limit, the sudden change in DDI energy results in the excitation of the scissors mode, a hallmark signature of superfluidity. The excitation protocol employed in this work leads to a distinct dependence of the scissors mode frequency on the effective DDI strength. We find that as the superfluidity of different phases evolves with changing effective DDI strength, each phase of the dipolar BEC exhibits a characteristic frequency response, demonstrating that this excitation approach provides a sensitive probe of superfluidity across different phases. Furthermore, while the superfluid phase exhibits an undamped scissors-mode oscillation, both the droplet and supersolid phases display damped oscillations. Interestingly, we find that the damping rate of the scissors oscillation provides a qualitative measure of the rigidity of the system. We further investigate the impact of the large angular deviation on the ground state phases. Our results reveal that, analogous to the elastic behavior of a solid, the deformation of individual droplets increases linearly with angular deviation up to a critical point. This elastic limit differs between the droplet and supersolid phases. Beyond this threshold, both the droplet and supersolid phases undergo permanent deformation and fail to recover their initial configurations even after the shearing stress is removed. In contrast, the superfluid state exhibits excitations of various undamped collective modes for large angular deviations of the polarization direction.\par 
	
	The subsequent material in this paper is arranged as follows. In Sec. \ref{secii}, we introduce the theoretical model in the form of an extended Gross-Pitaevskii equation (eGPE) governing the dynamics of the dipolar BEC. We describe the dynamics of different phases of dipolar BECs under the influence of a sudden change in polarization direction in Sec. \ref{seciii}. To this end, Subsection \ref{seciiia} focuses on the effects of small angular deviations in polarization, while Subsection \ref{seciiib} examines the impact of significant angular deviations. In Subsection \ref{seciiic}, we discuss the consequences of a finite-time linear ramp of the polarization direction as opposed to the instantaneous quenches of the previous subsections. We summarize our results and provide concluding remarks in Sec. \ref{seciv}. Appendix \ref{A} is devoted to the sum rule estimation of scissors mode frequency and Appendix \ref{B} discusses the effects of a double quench of the polarization direction.

	\begin{figure}[tb!]
		\centering
		\includegraphics[width=0.47\textwidth]{new_schematic_2d.pdf}
		\caption{Schematic illustration of the dynamical protocol to probe the signature of superfluidity and shear rigidity in a dipolar BEC. (a) Initial ground state configuration of a DDI dominated phase (SS or droplet) under an external magnetic field along the $z$-axis as represented by the gray dashed arrow [i.e., $\alpha(t)=0$ in Eq. \eqref{eq:poldirection}]. For the SF phase, a larger number of dipoles reside in the $z=0$ plane, leading to predominantly repulsive DDI. (b) Individual droplet with dipoles initially aligned along the $z$-axis, experience a shearing force (brown dashed arrow) following a sudden angular deviation $\alpha(t)=\Delta \theta$ at $t>0$. The initial and final polarization directions of the external magnetic field are indicated by the gray and violet dashed arrows, respectively.\label{fig:schematic}}
	\end{figure}
	
	\section{Model}\label{secii}
	We consider a BEC of dipolar atoms with a large magnetic dipole moment $\mu_m$ trapped in a three-dimensional cylindrically symmetric pancake-shaped harmonic potential $V_t(\vb{r})=\frac{1}{2}m[\omega_{\perp}^2(x^2+y^2)+\omega_z^2z^2]$, with $\omega_{\perp}=\omega_x=\omega_y<\omega_z$, where $\omega_{\perp}$ $(\omega_z)$ is the transverse (axial) trapping frequency and $m$ is the atomic mass. The atoms are polarized by a uniform external magnetic field whose direction lies in the $x$-$z$ plane and is given by 
	\begin{align}
		\vu{e}(t)=\vu{e}_x\sin{\alpha(t)}+\vu{e}_z\cos{\alpha(t)} \label{eq:poldirection},
	\end{align}
	with $\vu{e}_x$ and $\vu{e}_z$ being the unit vector along the $x$ and $z$-direction, respectively and $\alpha$ denotes the tilt angle with respect to the $z$-axis [see the schematic Fig. \ref{fig:schematic}(a) for $\alpha=0$ and Fig. \ref{fig:schematic}(b) for $\alpha=\Delta\theta$]. Here, we introduce a time dependence in the polarization to foreshadow the dynamics explored in the rest of the paper, where the polarization direction is modified by varying the tilt angle. In the ultracold temperature regime, the interaction between the particles can be modeled as a two-body pseudo-potential of the following form,
	\begin{equation}
		V_{\rm int}(\vb{r},t)=\frac{4\pi\hbar^2a_s}{m}\delta(\vb{r})+\frac{\mu_0\mu_m^2}{4\pi}\frac{1-3[\vu{e}(t)\cdot \vu{r}]^2}{r^3}.\label{v_int}
	\end{equation}
	The first term in Eq.~\eqref{v_int}, represents the isotropic and short-range contact interaction, characterized by the tunable $s$-wave scattering length $a_s$, and the second term describes the anisotropic and long-range DDI, where $\mu_0$ is the vacuum permeability and $\mu_m$ denotes the magnetic dipole moment. At such ultracold temperatures, the entire BEC system can be characterized by a single macroscopic wave function $\psi (\vb{r},t)$, also called the condensate order parameter, whose temporal evolution is governed by the extended Gross-Pitaevskii equation (eGPE) 
	\cite{chomaz_QuantumFluctuationDriven_2016, wachtler_Quantum_2016, halder_Control_2022}
	\begin{align}
		i\hbar\pdv{\psi(\vb{r},t)}{t}=&\Big[-\frac{\hbar^2}{2m}\nabla^2+V_t(\vb{r})\nonumber\\&+\int \dd\vb{r}' V_{\rm int}(\vb{r}-\vb{r}',t)\abs{\psi(\bf{r}^{\prime},t)}^2\nonumber\\&+\gamma(\epsilon_{\rm dd})\abs{\psi(\vb{r},t)}^3\Big]\psi(\vb{r},t).\label{egpe}
	\end{align}
	The last term in Eq.~\eqref{egpe}, takes into account the effect of quantum fluctuations in the form of dipolar Lee-Huang-Yang (LHY) correction with the coefficient $\gamma (\epsilon_{\rm dd})=(128\sqrt{\pi}\hbar^2a_s^{5/2}/3m) \left(1+\frac{3}{2}\epsilon_{\rm dd}^2\right)$ \cite{lima_Quantum_2011, lima_Meanfield_2012, petrov_Quantum_2015, schutzhold_MEANFIELD_2012, bisset_groundstate_2016}, where the dimensionless parameter $\epsilon_{\rm dd}=a_{\rm dd}/a_s$ quantifies the strength of DDI relative to the contact interaction with $a_{\rm dd}=\mu_0\mu_m^2 m/12\pi\hbar^2$ being the dipolar length. The LHY correction is essential for stabilizing the dipolar condensate against the mean-field driven collapse and allows the formation of exotic quantum droplet and supersolid phases. The condensate order parameter is normalized to the total number of atoms in the condensate \rm{i.e.} $N=\int d\textbf{r}\abs{\psi(\textbf{r})}^2$.\par 
	
	As we will discuss in forthcoming sections, our central results are valid in general for any dipolar BEC system at sufficiently low temperature to justify a mean-field description. Nevertheless, in order to illustrate our results, in this study we consider $N=6\times10^4$ number of $^{164}$Dy atoms confined in a quasi-2D, cylindrically symmetric harmonic potential with trapping frequencies $(\omega_{\perp},\omega_z)=2\pi\times(45,133)$Hz. The $^{164}$Dy atoms possess a magnetic dipole moment $\mu_m=9.93\mu_B$ which sets the dipolar length as $a_{\rm dd}=130.8a_0$, where $\mu_B$ and $a_0$ are the Bohr magneton and Bohr radius, respectively. The dipolar condensate exhibits either superfluid, supersolid, or droplet phases depending on the values of the dimensionless ratio $\epsilon_{\rm dd}$, which can be experimentally controlled by tuning the $s$-wave scattering length $a_s$ through the Feshbach resonance technique \cite{bottcher_Transient_2019, chomaz_LongLived_2019, tanzi_Observation_2019}. With the considered atom number, trapping parameter and a static magnetic field along the axial direction of the trap [i.e., along the $z$-axis with $\alpha(t) = 0$ as depicted in Fig. \ref{fig:schematic} (a)], the imaginary-time evolution of the eGPE~\eqref{egpe} yields the following ground state phases: a superfluid (SF) phase for $\epsilon_{\rm dd}<1.42$, a supersolid (SS) phase in the interval $1.42\leq\epsilon_{\rm dd}\leq1.5$, a multiple droplet (MD) phases for $1.5<\epsilon_{\rm dd}<1.7$ and a single droplet (SD) phase for $\epsilon_{\rm dd}>1.7$. \par

	
	\section{Dynamics under the influence of sudden angular deviation of polarization direction}\label{seciii}

	We begin our calculations by obtaining the ground states corresponding to a superfluid with $a_s=120a_0$ $(\epsilon_{\rm{dd}}=1.09)$, a supersolid with $a_s=90a_0$ $(\epsilon_{\rm{dd}}=1.45)$ and a droplet with $a_s=70a_0$ $(\epsilon_{\rm{dd}}=1.86)$ confined in the pancake-shaped trap in the presence of a static magnetic field along the axial ($z$-) direction of the trap [see the schematic in Fig. \ref{fig:schematic}(a)] by the imaginary time evolution of the eGPE \eqref{egpe}. Subsequently, we generate dynamics by suddenly switching the polarization direction of the external magnetic field $\vu{e}$ from its initial orientation by an angular deviation $\Delta \theta$ \rm{i.e.} our protocol can be summarized as (see the schematic in Fig. \ref{fig:schematic})
	\begin{align}
		\alpha(t) = \begin{cases}
			0, & t<0, \\
			\Delta \theta, & t>0 .
		\end{cases}
	\end{align}
	Due to the sudden change in polarization, the condensate no longer remains in the ground state of the new configuration and undergoes a nonequilibrium dynamics. A qualitative picture of the dynamics one expects can be obtained by the following argument. To minimize the DDI energy, all the atoms will try to align along the altered polarization direction. The cumulative impact of this on the system, arising from the dipolar interactions between the atoms, manifests as a shearing stress proportional to $\Delta\theta$ [depicted by the brown dashed arrow in Fig. \ref{fig:schematic}(b)]. As we show in detail below, the response of the dipolar BEC to such a shearing stress depends on its ground state phase (SF, SS, or droplet), and can serve as a distinctive signature that showcases the characteristic properties of each phase.
	
	\begin{figure*}[tb!]
		\centering
		\includegraphics[width=\textwidth]{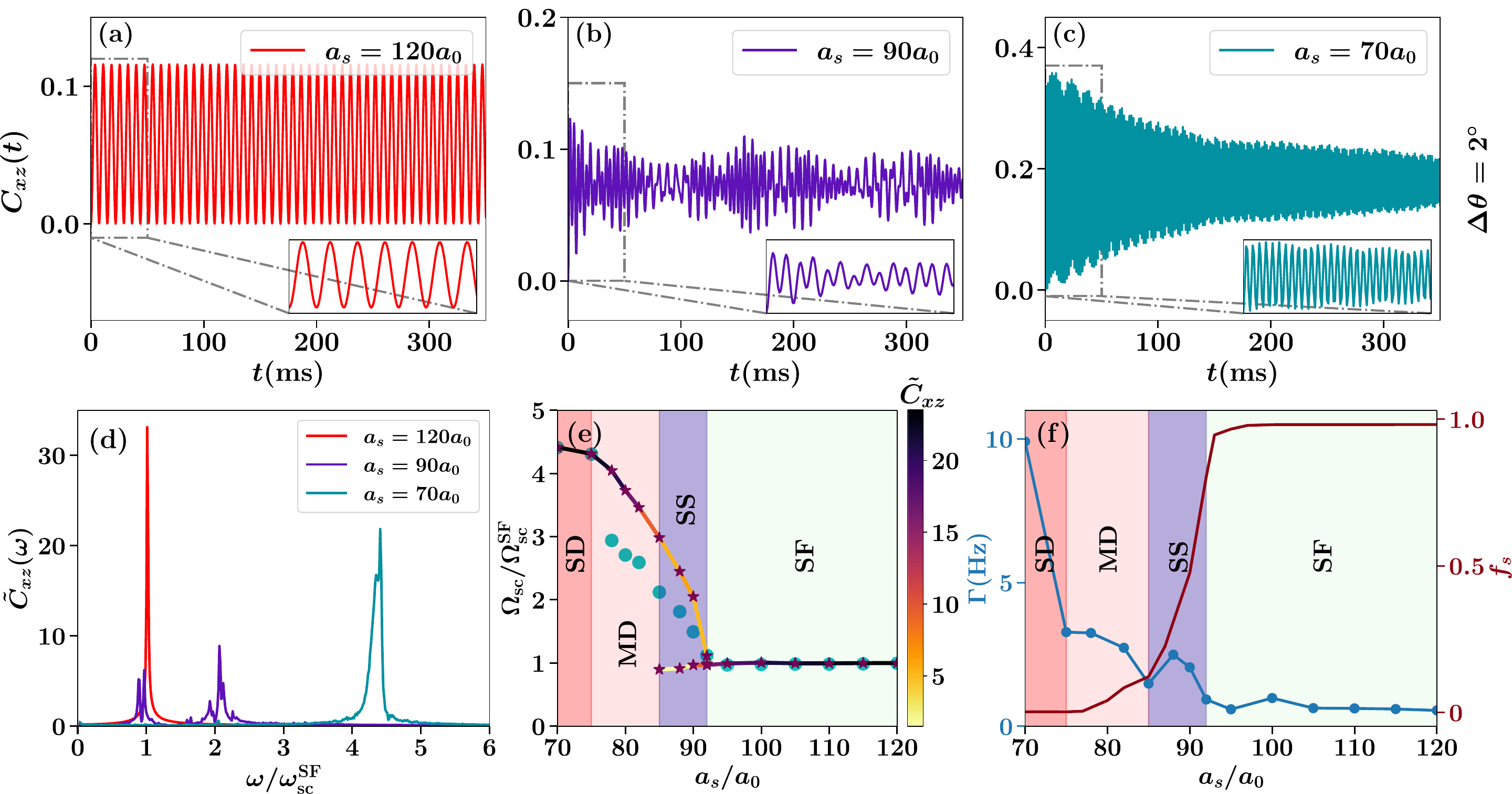}
		\caption{The panels $\rm{(a-c)}$ demonstrate the temporal evolution of the scissors mode in the $x$-$z$ plane for (a) the SF ($a_s=120a_0$), (b) SS ($a_s=90a_0$) and (c) the SD ($a_s=70a_0$) state, respectively, following a sudden change in polarization direction by $\Delta \theta=2^{\circ}$. The insets in $\rm{(a- c)}$ display an enlarged view of $C_{xz}(t)$ within the time interval from $t =0-50$ ms. Panel (d) depicts the corresponding frequency spectrum $\tilde{C}_{xz}(\omega)$ obtained by the Fourier transformation of the signal $C_{xz}(t)$ shown in $\rm{(a-c)}$. Panel (e) shows the scissors mode frequency as a function of $a_s$ ($\Omega_{sc}^{\rm SF} = 2\pi \times 135 ~\rm{Hz}$). Brown star-shaped markers correspond to the prominent frequency of the signal $C_{xz} (t)$ and the teal colored circles correspond to the frequency estimated by the sum rule (see Appendix \ref{A}). Panel (f) illustrates the variation of the superfluid fraction $f_s$ and the HWHM ($\Gamma$) of the dominant peak in the Fourier spectrum $\tilde{C}_{xz}(\omega)$, which quantifies the damping rate of the corresponding state. The background color shading in $\rm{(e)–(f)}$ delineates the different phase domains obtained from the imaginary-time evolutions of the eGPE.} \label{fig:1}
	\end{figure*}	
	\subsection{Impact of small angular deviation in polarization direction}\label{seciiia}
	We first consider the situation of sudden change of polarization direction by a small angle, such that the resulting dynamics is still within the linear response regime for the BEC. In such a limit, the change in the polarization direction of the atomic dipoles produces a small shearing stress on the condensate and thereby generates the $y$-component of the angular momentum leading to the excitation of scissors mode oscillation in the $x$-$z$ plane \cite{ferrier_barbut_scissors_2018}. The dynamics associated with the scissors mode oscillation of the condensate can be monitored by the time evolution of the expectation value
	\begin{equation}
		C_{xz}(t) = \expval{xz} = \int \dd{\vb{r}} \,\, x z \vert \psi(\mathbf{r},t) \vert^2 \label{eq:C_xz}
	\end{equation}
	as illustrated in Figs. \ref{fig:1}(a)-\ref{fig:1}(c). In the SF phase, the condensate undergoes an undamped oscillation as shown in Fig.~\ref{fig:1}(a). However, in the case of the SS and droplet phases, atoms are displaced from their initial equilibrium orientation and relax into a new strained equilibrium orientation following a damped scissors mode oscillation to minimize the DDI as long as the shearing stress governed by the polarization alteration is present [see Figs. \ref{fig:1}(b) and \ref{fig:1}(c)]. Note that we have shown the result for a SD in Fig.~\ref{fig:1}(c) but the result remains the same for the MD case too. In this small angular deviation limit of the polarization direction, removing the stress by restoring the orientation to the initial polarization direction, the atoms in both the droplet (SD/MD) and SS states realign with the initial polarization direction, again through damped scissors-mode oscillations (see Fig.~\ref{fig:8} in  Appendix \ref{B}). This behavior is analogous to the elastic response of a solid within its elastic limit.\par
	To extract the scissors mode frequency, denoted as $\Omega_{\mathrm{sc}}$, we perform a Fourier transform of the expectation value $C_{xz}(t)$ defined in Eq.~\eqref{eq:C_xz}. The dominant peaks in the resulting Fourier spectrum $\tilde{C}_{xz}(\omega)$, shown in Fig.~\ref{fig:1}(d), can be used to read off the characteristic frequency of the scissors mode. In the SF phase, the condensate exhibits a single sharp peak in the frequency spectrum. We also found that with an increase in the angular deviation of the polarization direction, the amplitude of the scissors-mode oscillation increases, but the frequency remains almost unchanged (not shown here). This suggests that although the scissors-mode excitation in the SF phase is triggered by the sudden change in the polarization angle, the resulting oscillation frequency is only weakly influenced by dipolar interactions and is instead primarily determined by the trap anisotropy (see Appendix \ref{A} for more details). However, as the dipolar interaction increases relative to the contact interaction, the condensate enters into a quasi-2D SS phase domain, where the frequency spectrum reveals the appearance of a new broadened scissors-mode excitation peak at a higher frequency in addition to the lower-frequency mode akin to that in the SF phase [see Fig.~\ref{fig:1}(d)]. Thus, the higher frequency mode is associated with the scissors-mode excitation of the droplet-like density peaks in the SS phase, while the lower frequency corresponds to the excitation arising from the background superfluid. As a consequence of this binary excitation, the temporal evolution of $C_{xz}(t)$ demonstrates a beating pattern [see Fig.~\ref{fig:1}(b)]. As the dipolar interaction is increased further to the SD phase, as shown in Fig.~\ref{fig:1} (d), only the high-frequency oscillatory modes persist. Moreover, as evident from Fig.~\ref{fig:1}(d), the corresponding spectral peak exhibits a substantial broadening in this regime, which stems from multiple closely spaced frequency components at higher frequencies. The latter arise from nonequilibrium quenches in the polarization direction, together with the subsequent dephasing among these modes due to nonlinear interactions, resulting in an effective damping of the scissors-mode oscillations. We also note that the decay of the scissors-mode amplitude does not represent true dissipative damping, since both the particle number and the total energy of the condensate remain conserved throughout the dynamics. A similar occurrence of multiple frequency components and broadened excitation peaks in the excitation spectrum has been reported previously, particularly in dipolar-interaction–dominated phases such as the SS and droplet phases when subjected to sudden trap rotation \cite{roccuzzo_Moment_2022, norcia_Can_2022}. In those studies, a similar damping of the scissors mode oscillations was also observed.\par
	
	In Fig.~\ref{fig:1}(e), we have depicted the variation of the scissors mode frequency with the $s$-wave scattering length, which demonstrates the bifurcation of $\Omega_{\rm{sc}}$ as the system undergoes the SF to SS phase transition. By further decreasing $a_s$, both the frequency and amplitude of the lower-frequency scissors mode gradually decrease and eventually vanish at the SS to droplet phase transition boundary, whereas the frequency of the higher-frequency droplet scissors mode continues to increase with increasing effective DDI strength.
	Although the characteristic low-frequency scissors-mode excitation associated with the SF phase vanishes in the MD states, the emergence of a two-dimensional crystalline droplet structure leads to a complex excitation spectrum. This spectrum features multiple lower-amplitude peaks along with a dominant broadened excitation peak in the high-frequency region. In Fig.~\ref{fig:1}(e), we only depict the prominent peaks for clarity.\par
	
	We note that the observed increase in the scissors-mode frequency stands in contrast to the conventional scissors-mode excitation protocol by applying a sudden trap rotation with angular momentum along the polarization direction of the external magnetic field. In that scenario, both theoretical and experimental studies in an axially elongated trap show that the scissors-mode frequency decreases as the system transitions from the SF phase to the droplet state (SD/ MD) via the intermediate SS phase \cite{roccuzzo_Rotating_2020, tanzi_Evidence_2021}. Indeed, as reported in \cite{norcia_Can_2022}, the scissors mode frequency excited in a similar way is independent of the DDI and remains nearly unchanged, stays close to the prediction for a purely SF phase and far away from the rigid body prediction for the isolated droplet crystal phase (SD and MD). This effect becomes significantly more pronounced in a two-dimensional trapping geometry. In contrast, as shown in Fig.~\ref{fig:1} (e), under the protocol employed in this work, the scissors-mode frequency displays a strong dependence on the effective DDI and varies significantly as the condensate evolves from the superfluid to the droplet phase.\par
	
	For further validation of our findings and obtaining qualitative insights into this behavior, we first note that a rigorous upper bound for the lowest scissors mode frequency of the form
	\begin{align}
		\hbar\omega_{sc}\leq\sqrt{\frac{m_1}{m_{-1}}},\label{omega_sc}
	\end{align}
	follows from linear response theory and sum rules (which is valid for small angular deviation of the polarization direction) \cite{pitaevskii_bose-einstein_2016}. Here $m_1 = \hbar^2 \int \dd\omega \, \omega S (\hat{L}_y,\omega)$ represents the energy-weighted moment and $m_{-1} = \int \dd\omega \, S(\hat{L}_y, \omega)/ \omega$ is known as the inverse energy-weighted moment of the dynamical structure factor $S(\hat{L}_y,\omega) = \sum_n\abs{\mel{n}{\hat{L}_y}{0}}^2\delta(\hbar\omega-\hbar\omega_n)$ (taking $T=0$ for simplicity), associated with the angular momentum operator $ \hat{L}_y=-i\hbar(z\partial_x-x\partial_z)$. Interestingly, the moment $m_1$ can be expressed in terms of a double commutator involving the Hamiltonian of the system as $m_1 = \frac{1}{2} \langle [\hat{L}_y, [\hat{H}, \hat{L}_y]] \rangle,$ where the expectation value is taken with respect to the equilibrium configuration of the system and $\hat{H}$ represents the many-body Hamiltonian of the dipolar BEC. Physically, $m_1$ serves as an effective restoring force for the scissor-mode oscillation \cite{ferrier_barbut_scissors_2018, norcia_Can_2022}. In our case, the nonlinear DDI between the atoms set by the external polarizing magnetic field combined with the trap anisotropy $(\omega_{\perp}=\omega_x=\omega_y \neq \omega_z)$ breaks the rotational symmetry of the condensate in the $x$-$z$ plane and contributes to the commutator $[\hat{H}, \hat{L}_y] \neq 0$. Under such conditions, the energy-weighted moment $m_1$ simplifies to (see Appendix \ref{A} for the detailed derivation)
	\begin{align}  
		m_1 &= \frac{1}{2} \left(\langle [\hat{L}_y, [\hat{V}_{\rm dd}, \hat{L}_y]] \rangle + \langle [\hat{L}_y, [\hat{V}_{t}, \hat{L}_y]] \rangle\right)\nonumber\\& =\frac{\hbar^2}{2} \left( \langle V_{\rm dd}^x \rangle -  \langle V_{\rm dd}^z \rangle +m (\omega_z^2-\omega_x^2)\langle x^2-z^2\rangle  \right),\label{m_1}
	\end{align}
	where $\langle V_{\rm dd}^\eta \rangle = \int \dd{\vb{r}} \dd{\vb{r}^\prime} \abs{\psi(\vb{r})}^2 V_{\rm dd}^\eta(\vb{r}-\vb{r}^\prime)\abs{\psi(\vb{r}^{\prime})}^2,$ 
	and $V_{\rm dd}^\eta(\vb{r}-\vb{r}^\prime) = \frac{\mu_0\mu_m^2}{4\pi\abs{\vb{r}-\vb{r'}}^3}\left[1-3\frac{(\eta-\eta')^2}{\abs{\vb{r}-\vb{r'}}^2}\right]$, and $\eta = x, y, z.$ We note that Eq.~\eqref{m_1} differs considerably from the one obtained for usual scissors-mode excitation in a dipolar (non-dipolar) BEC when rotated about the polarization direction of the external magnetic field (axial direction of the trap, \rm{i.e.,} the $z$-axis) \cite{roccuzzo_Rotating_2020, tanzi_Evidence_2021, roccuzzo_Moment_2022, young-s._Dipolemode_2023, guery-odelin_Scissors_1999, marago_Observation_2000}. In that case, the commutator $[\hat{V}_{\rm dd}, \hat{L}_z]$ vanishes and the energy weighted moment $m_1$ or the effective restoring force is solely governed by the trap anisotropy. Since elasticity fundamentally arises from interparticle interactions, no elastic response of the condensate is involved in the scissors mode oscillation. In contrast, in our case, not only the trap but also the DDI between the atoms plays a crucial role in shaping the restoring force [see Eq.~\eqref{m_1}]. Consequently, following our dynamical protocol, the excited scissors-mode oscillation bears a clear imprint of the elastic response, which is determined by the anisotropy in the DDI, \rm{i.e.} $\langle V_{\rm dd}^x \rangle - \langle V_{\rm dd}^z \rangle$, between the particles of the condensate in the $x$–$z$ plane\footnote{A similar method is employed in \cite{ferrier_barbut_scissors_2018}, where an oscillating magnetic field is applied to excite the scissors mode in the SD phase. In this setup, only the DDI contributes to $m_1$, while the trap symmetry prevents the trapping potential from playing any role in determining the effective restoring force.}. \par
	
	On the other hand, the inverse energy weighted moment $m_{-1}$ is related to the moment of inertia of the condensate $\Theta$ through the relation $m_{-1}=\Theta/2$. We estimate $\Theta$ by evaluating the static response of the system to a small rotational perturbation of the form $-\Omega \hat{L}_y$. The moment of inertia is then obtained as $\Theta=\lim_{\Omega\to 0}\langle \hat{L}_y\rangle/\Omega$. The response to this small static perturbation, or equivalently the moment of inertia varies across different phases of the dipolar BEC. More specifically, the moment of inertia $\Theta$ depends on the superfluid fraction $f_s$ or the nonclassical rotational inertia fraction of the condensate and can be expressed as \cite{leggett_Superfluid_1998, roccuzzo_Rotating_2020, tanzi_Evidence_2021}
	\begin{equation}
		\Theta = 2m_{-1}=(1-f_s)\Theta_{\rm{rig}}+f_s \Theta_{\rm{SF}},\label{m_-1}
	\end{equation}
	where $\Theta_{\rm{rig}}=m\langle x^2+z^2\rangle$ is the classical rigid body value of the moment of inertia about the $y$-axis, $\Theta_{\rm{SF}}=\beta^2\Theta_{\rm{rig}}$ is the moment of inertia corresponding to that of a superfluid and $\beta=\langle x^2-z^2\rangle/\langle x^2+z^2\rangle$ is a geometrical factor quantifying the degree of deformation of the condensate. The SF and droplet states correspond to a superfluid fraction of $f_s\approx 1$ and $f_s\approx 0$, respectively, while the SS state exhibits an intermediate value between $0 < f_s<1$. Using the ground state solution in the presence of a static magnetic field along the axial direction, we determine the values of $m_{1}$ and $m_{-1}$ and subsequently use the same in Eq.~\eqref{omega_sc} to obtain the upper bound of the scissors mode frequency set by the sum rule and take it as an estimate \emph{i.e.} $\Omega_{\mathrm{sc}} \approx \omega_{\mathrm{sc}}$. We find that in the small angular deviation limit ($\Delta\theta=2^{\circ}$) and for the case of SF and SD, where the scissors mode excitation is peaked at a single frequency, this sum rule estimate matches well with the $\Omega_{\mathrm{sc}}$ extracted from the real-time simulation of the eGPE [Eq.~\eqref{egpe}] as shown in Fig.~\ref{fig:1} (e) by the teal circular markers. In the case of the SS and MD states, due to the excitation of multiple frequency peaks\footnote{In Fig.~\ref{fig:1}(e), we present only the prominent frequency peaks (brown star-shaped markers). In the MD states, we observe that, in addition to the prominent high-frequency scissors-mode peak, several low-frequency excitations with very small amplitudes are also present, even for a very small angular deviation of $\Delta\theta=2^{\circ}$. However, these minor peaks are not displayed in Fig.~\ref{fig:1}(e) for clarity. The sum-rule estimate (cyan colored markers) provides an upper bound for the lowest-frequency mode.}, $\omega_{\mathrm{sc}}$ from Eq.~\eqref{omega_sc} does not agree with the different frequencies extracted from the numerical simulation of eGPE, but nonetheless provides the upper bound for the lowest frequency scissors mode excitation. Most importantly, we find that although $m_{-1}$ increases with decreasing $f_s$ (or $a_s$, see Eq. \eqref{m_-1}), the increase in $m_1$ is more significant due to the increasing dipolar anisotropy. As a result, the scissors-mode frequency increases as the system transitions from the SF to the droplet phase (SD/MD). This growth in the scissors mode frequency contrasts with earlier works \cite{roccuzzo_Rotating_2020, tanzi_Evidence_2021}, where the scissors mode was excited by a sudden rotation of the trap about the polarization axis. In that scenario, $m_1$ is solely governed by the trap anisotropy and remains fixed for a given trapping configuration. On the other hand, $m_{-1}$ increases as  $f_s$ (or $a_s$) decreases, which leads to a reduction of the scissors-mode frequency.\par

	Moreover, we find that the effective damping rate of the signal $C_{xz}(t)$ is directly correlated with the width of the dominant frequency peak in the spectrum. To gain further insight, we fit the dominant frequency peak in the numerically extracted spectrum $\tilde{C}_{xz}(\omega)$ with a Lorentzian function of the form
	\begin{equation}
		\tilde{C}_{\rm{fit}}(\omega)=\frac{A\Gamma^2}{(\omega-\Omega_0)^2+\Gamma^2},
	\end{equation}
	where $A$ denotes the amplitude of the dominant frequency peak at a frequency $\Omega_0$ and $\Gamma$ denotes the half-width at half-maximum (HWHM). A larger value of $\Gamma$ signifies stronger damping, which originates from the attractive component of the DDI as evident from its monotonic increase with decreasing $a_s/a_0$ in Figure \ref{fig:1} (f). This attractive DDI tends to align the dipoles along the modified polarization axis, in competition with the scissors-mode oscillation that drives the angular oscillation of the condensate. As the effective DDI increases over the contact interaction, the scissors-mode oscillation becomes increasingly damped and aligns rapidly with the modified polarization direction. Moreover, we also observe from Fig.~\ref{fig:1} (f) that there is a direct correlation between the decreasing superfluid fraction $f_s$ and increasing $\Gamma$ (as $a_s/a_0$ is decreased), demonstrating that $\Gamma$ can also serve as a qualitative measure of the rigidity of the system. 
	
	\begin{figure*}[tb!]
		\centering
		\includegraphics[width=\textwidth]{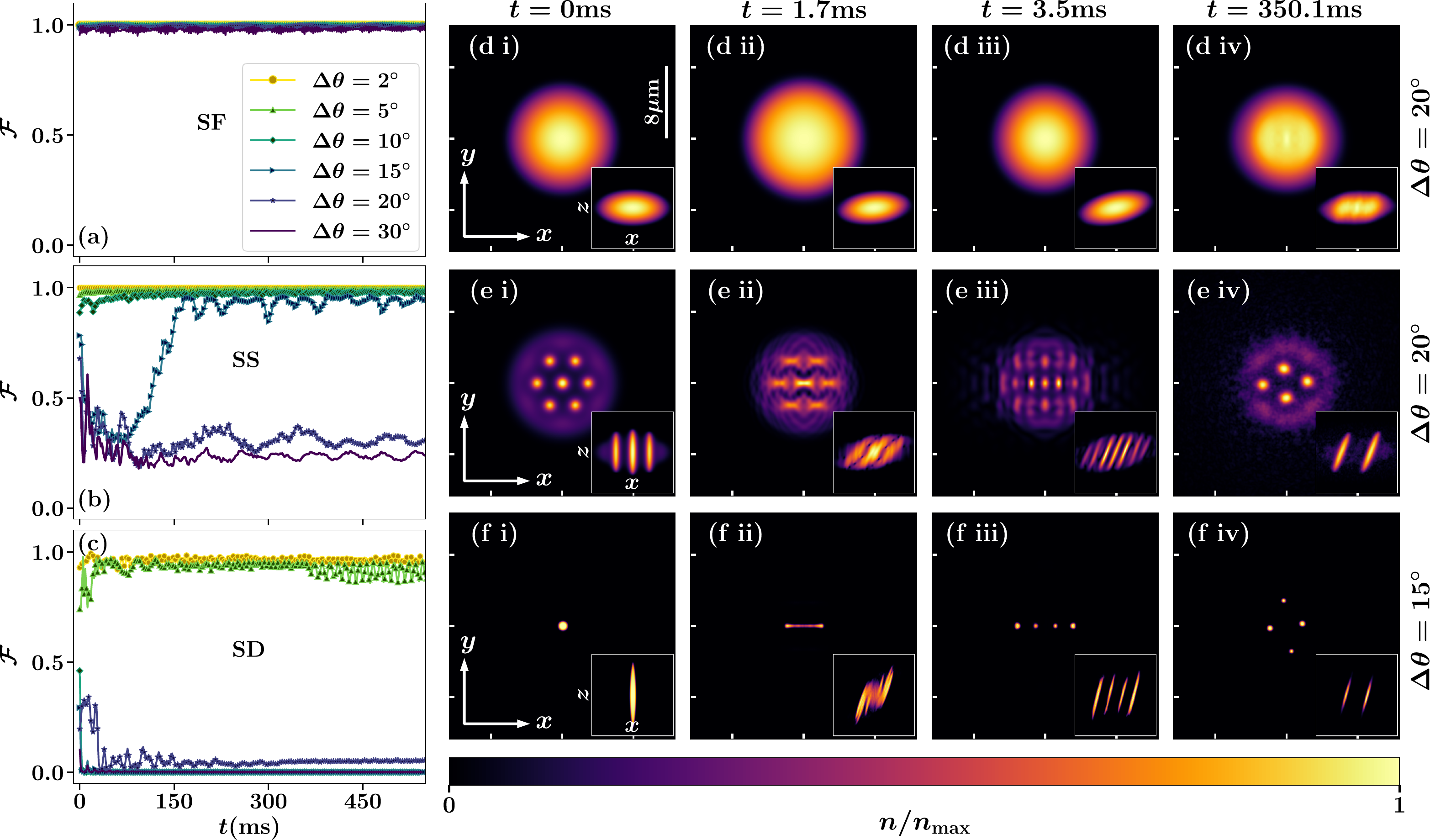}
		\caption{Time evolution of the fidelity of (a) SF, (b) SS and (c) SD state under the influence of the sudden change in polarization direction. The different colored lines represent different angular deviations, as shown in the legend. $\rm{(d i-d iv)}$ Represent the density profiles $n(x,y,z=0,t)$ of the SF state under a $\Delta \theta = 20^\circ$ angular deviation, resulting in $\mathcal{F} = 1$, thereby no structural change occurs. $\rm{(e i-e iv)}$ Illustrate the density profile of the SS state at different times after the sudden $\Delta \theta = 20^\circ$ change in polarization direction, resulting in a structural deformation of the crystalline structure, with $0<F < 1$. $\rm{(f i-f iv)}$ Depict the density profile of the SD state at different time intervals following a sudden change in the polarization direction of $\Delta \theta = 15^\circ$, causing permanent deformation ($\mathcal{F} = 0$) and transition to the MD state. Insets in $\rm{(d-f)}$ demonstrate the corresponding density distribution in the $y=0$ plane.}\label{fig:2}
	\end{figure*}	
	\subsection{Impact of large angular deviation in polarization direction}\label{seciiib}
	Let us now consider the situation where, starting from the ground state, a sudden, large angular deviation in the polarization direction is induced. As we will demonstrate in this section, this leads to exotic non-equilibrium dynamics, including a transition in the mechanical response from elastic to plastic-like behavior of the system when the dipolar interactions dominate. As the first dynamical variable of interest, we consider the deviation from the ground state $\ket{\Psi_{\Delta \theta}}$ prepared in the presence of a static tilted magnetic field \rm{i.e.} the ground state of Eq.~\eqref{egpe} with $\alpha(t) = \Delta \theta$, and the dynamically evolved state $\ket{\psi(t})$ under a sudden change in the polarization direction by the same amount quantified by the fidelity $\mathcal{F}(t)=\abs{\braket{\Psi_{\Delta \theta}}{\psi(t)}}^2$. A fidelity of $\mathcal{F}=1$ indicates that the states are identical, whereas $\mathcal{F}=0$ corresponds to orthogonal states that are completely distinct. The intermediate values $0<\mathcal{F}<1$ reflect partial overlap, implying that the states share some similarity but remain different from the ground state prepared under the static condition.\par
	Focusing first on the SF phase we note that as we increase $\Delta \theta$, the condensate exhibits various low-lying excited modes along with the scissors mode oscillation. While the amplitude of these low-lying modes increases with the angular deviation, regardless of the value of $\Delta \theta$ the fidelity of the SF phase remains consistently close to $\mathcal{F}=1$ as shown in Fig.~\ref{fig:2} (a). The fidelity demonstrates slight oscillations due to the excitation of the various low-lying modes pictured via the density plots Fig.~\ref{fig:2}(d i - d iv). Consequently, the SF phase exhibits a non-elastic response even when there is a significant angular deviation in the polarization direction.\par 
	A completely different scenario unfolds in the case of the SD and SS states. For the SS and the SD state, beyond a critical angular deviation $\Delta \theta_c$, the fidelity drops from $1$, indicating destabilization of the initial state and formation of a new metastable state [see Fig.~\ref{fig:2} (b)-(c)]. Focusing first on the SS phase, we can understand the behavior of the fidelity in Fig.~\ref{fig:2} (b) as follows. As $\Delta \theta$ increases the shear stress on each droplet of the SS increases, leading to a deformation of the shape of the droplets. Owing to the presence of the superfluid background, the dense droplets can temporarily deform under shear stress but subsequently relax back to their initial configuration after a certain time period. Consequently, the fidelity exhibits a backbending behavior dropping below unity and later returning to unity again within the elastic regime of the SS \rm{i.e.} for $\Delta \theta < \Delta \theta_c$. However, upon further increasing the angular deviation, a threshold is reached at $\Delta \theta=\Delta \theta_c$ beyond which the SS can no longer sustain the shear stress and undergoes permanent deformation. The number of the droplets as well as the lattice structure undergo a transition from a triangular lattice to a rectangular lattice formation as shown in Figs. \ref{fig:2} (e i- e iv). Interestingly, in spite of this structural transition evident in the density plots, due to the presence of a non-vanishing superfluid fraction, there is always a small finite overlap between the dynamically evolved state and the ground state prepared at that tilt angle. This results in fidelity $0<\mathcal{F}<1$, even when $\Delta \theta> \Delta \theta_c$ [see Fig.~\ref{fig:2} (b)] for the SS state.\par
	Turning our attention to the droplet phase, we find that the single droplet exhibits a damped scissors mode oscillation at small values of $\Delta \theta$, keeping the fidelity $\mathcal{F}\sim 1$ at all times as shown in Fig.~\ref{fig:2} (c). As the shearing stress increases with increasing $\Delta\theta$ beyond the elastic limit, the fidelity suddenly drops to zero and the SD state undergoes an elastic to plastic phase transition [see Fig.~\ref{fig:2} (c)]. Beyond the critical angle $\Delta \theta> \Delta \theta_c$, the SD initially breaks into an array of MD along the $x$-axis [see Fig.~\ref{fig:2} (f iii)]. Owing to the symmetric trapping configuration and the inter-droplet interactions, this 1D chain subsequently evolves into a two-dimensional rectangular lattice structure in the $x–y$ plane as shown in Fig.~\ref{fig:2} (f iv). We note that the elastic to plastic transition evidenced for the SD and SS phases is irreversible and leads to a permanent deformation of the state in the sense that restoring the polarization direction to its original orientation does not cause the system to recover the initial configuration as shown in Fig.~\ref{fig:9} in Appendix B. By examining the behavior of the fidelity for different $\Delta \theta$, we find that the critical angle amounts to $\Delta \theta_c \sim 15^{\circ}$ for the SS state and $\Delta \theta_c \lesssim 10^{\circ}$ for the SD. Thus, the critical angular deviation $\Delta \theta_c$ needed for the elastic to plastic phase transition is lower for the SD than in the SS state. This suggests that the SD state is more rigid than the SS state. The distinction can be attributed to the presence of a superfluid background and global phase coherence in the SS state. The superfluid background allows particles to transfer between droplets, thereby reducing the effective stress on each droplet, which in turn increases the elastic regime for the SS state. \par
	\begin{figure}[tb!]
		\centering
		\includegraphics[width=0.47\textwidth]{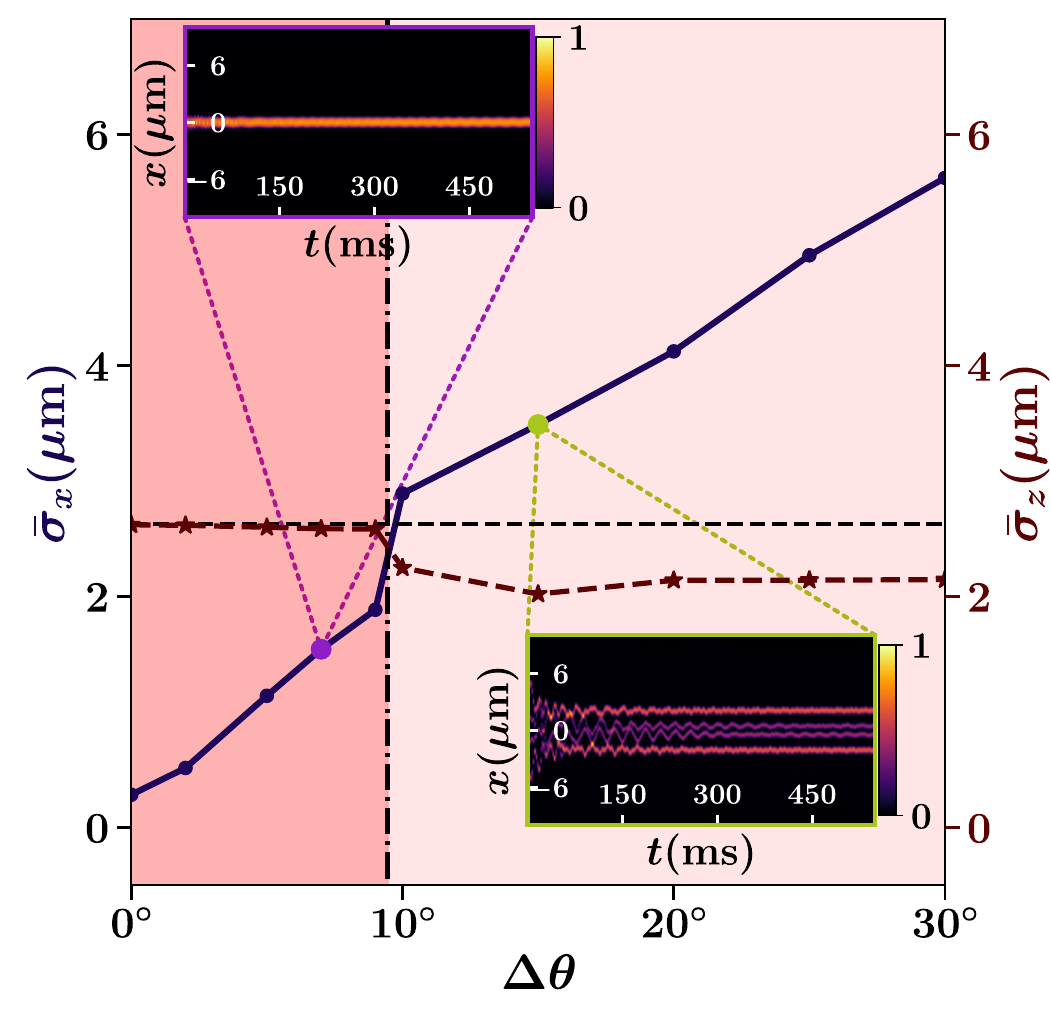}
		\caption{Variation of the average rms width $\bar{\sigma}_x$ along the $x$-axis (solid dark blue line with circular markers) and the average rms width $\bar{\sigma}_z$ along the axial direction of the condensate initially in the SD phase (brown dashed line with star shape markers) as a function of the $\Delta \theta$ following the sudden change in polarization direction. The horizontal black dashed line indicates the value of axial width of the ground state in the presence of a static magnetic field along the $z$-axis. The shaded regions indicate the elastic ($\bar{\sigma}_x <\bar{\sigma}_z$) and plastic ($\bar{\sigma}_x >\bar{\sigma}_z$) regimes of the SD state, with the vertical dashed line marking the transition point from elastic to plastic phase domain. The insets demonstrate the temporal evolution of the integrated density profile $n (x)/n_{\rm max}$ corresponding to the highlighted points in the main plot.\label{fig:avg_correlation_length}}
	\end{figure}
	To characterize the elastic to plastic transition in more detail and to further probe the development of permanent deformation during the dynamical process, we consider the integrated density along the $x$-direction  $n(x,t) = \int \dd{y} \dd{z} \vert \psi(\mathbf{r},t) \vert^2,$ as well as the time-averaged root mean square (rms) width along the $x$-axis ($z$-axis) $\bar{\sigma}_x$ ($\bar{\sigma}_z$). In Fig.~\ref{fig:avg_correlation_length}, we show the behavior of $\bar{\sigma}_x$ and $\bar{\sigma}_z$ for the SD case for different values of $\Delta \theta$.
	At small angular deviations, due to the magnetostriction effect, in order to minimize the DDI energy, the axial width of the SD state $\bar{\sigma}_z$ remains larger than the transverse width $\bar{\sigma}_x$. In this regime, $\bar{\sigma}_x$ increases linearly with $\Delta \theta$ and $\bar{\sigma}_z$ is essentially constant. Moreover, reversing the polarization direction back to its initial value allows the SD state to retrace the same loading path and return to its initial value, indicating the elastic nature of this regime. As $\Delta \theta$ is increased to larger values and crosses the critical value $\Delta \theta_c$, the SD state becomes dynamically unstable under the influence of the large shearing stress and as the insets in Fig.~\ref{fig:avg_correlation_length} show, $n(x,t)$ goes from a single peaked structure (SD) to multiple peaks (MD). The emergence and the persistence of these additional peaks during the subsequent time evolution signal the onset of a metastable periodic density modulation. This process is accompanied by a sudden increase in $\bar{\sigma}_x$ and an associated sudden decrease in $\bar{\sigma}_z$. Indeed the interpolated crossing between $\bar{\sigma}_x$ and $\bar{\sigma}_z$ marked with a dashed vertical line in Fig.~\ref{fig:avg_correlation_length} can be used to obtain the value of $\Delta \theta_c \lesssim 10 ^{\circ}$ for the SD regime. Note that resetting the polarization angle to the initial value after the initial quench for $\Delta \theta > \Delta \theta_c$ will not restore the initial state. Thus such an angular deviation of the polarization direction results in a permanent and irreversible deformation of the SD state (see Appendix \ref{B}).\par

	\begin{figure}[htb!]
		\centering
		\includegraphics[width=0.48\textwidth]{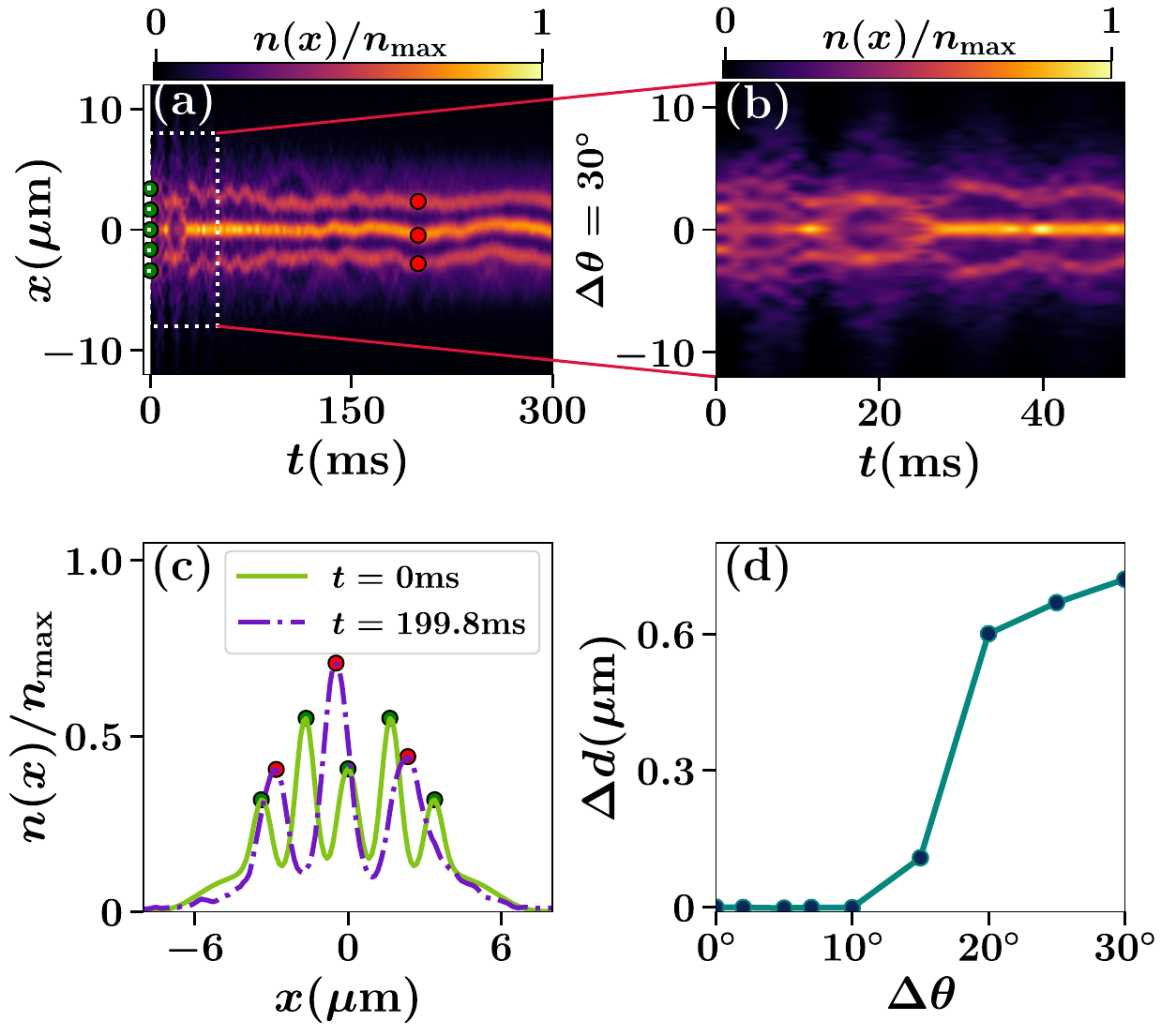}
		\caption{(a) Temporal evolution of the integrated density profile $n (x)/n_{\rm max}$ for the SS subjected to a sudden angular deviation of $\Delta \theta=30^{\circ}$ in the polarization direction. Panel (b) displays an enlarged view of the temporal evolution of the density profile within the interval $t=0-50~\rm{ms}$. Panel (c) demonstrates the spatial density distribution $n (x)/n_{\rm max}$ at time $t=0 ~\rm{ms}$ and $t=199.8 ~\rm{ms}$. The corresponding correlated density peaks are indicated by green and red circular markers, consistent with those shown in panel (a). (d) Variation of the density peak separation, measured relative to its initial value, as a function of the angular deviation $\Delta \theta$.}\label{fig:ss_elastic_plastic}
	\end{figure}
	In the SS state, due to the periodic density modulation, the integrated density profile $n(x,t=0)$ in the initial ground state (before the polarization direction quench) itself exhibits multiple peaks as shown in Figs. \ref{fig:ss_elastic_plastic}(a) and \ref{fig:ss_elastic_plastic}(c). Under the influence of the sudden change in polarization direction beyond the critical threshold ($\Delta \theta> \Delta \theta_c$), both the number of peaks and their positions shift from the initial configuration and persist throughout the subsequent dynamics as shown in Fig.~\ref{fig:ss_elastic_plastic} (a-c). Since a simple measure such as the width of the condensate is not able to sharply capture this transition, we consider an alternative measure as follows. Let $N_j$ be the number of correlated peaks in $n(x,t_j)$ at a given time instant $t_j$, located at positions $x_i^j$ with $i = 1,2,\dots,N_j$. The mean separation between these peaks at the time instant $t_j$ can be expressed as $d(t_j)=\sum_{i=1}^{N_j-1}(x_{i+1}^j-x_i^j)/(N_j-1)$. To quantify the cumulative effect of the change in the density profile at the transition as measured by the emergence of new peaks in $n(x,t)$, we define the deviation $\Delta d=\overline{d(t_j)}-d(0)$, where $\overline{d(t_j)}$ denotes the time-averaged separation between the correlated peaks and $d(0)$ corresponds to the initial spacing. Fig.~\ref{fig:ss_elastic_plastic} (d) illustrates how this deviation $\Delta d$ varies with $\Delta \theta$. We clearly see a sudden growth of the deviation $\Delta d$ beyond a critical value of $\Delta \theta$ and can estimate the same from Fig.~\ref{fig:ss_elastic_plastic} (d) as $\Delta \theta_c \approx 15 ^{\circ}$. As in the SD case, polarization quenches with $\Delta \theta >\Delta \theta_c$ lead to permanent changes in the crystalline structure of the SS state. While we do not discuss the behavior of $n(x,t)$ for the SF state in detail here (since it is not remarkable), we note that for all times we find a single broad density peak at $x=0$ and no new correlated density peaks centered at positions $x\neq0$, regardless of the value of $\Delta \theta$. Thus, unlike the SD and SS cases, the SF state does not exhibit an elastic to plastic phase transition when subjected to the shear stress caused by sudden changes of the polarization direction.

	\begin{figure}[tb!]
		\centering
		\includegraphics[width=0.42\textwidth]{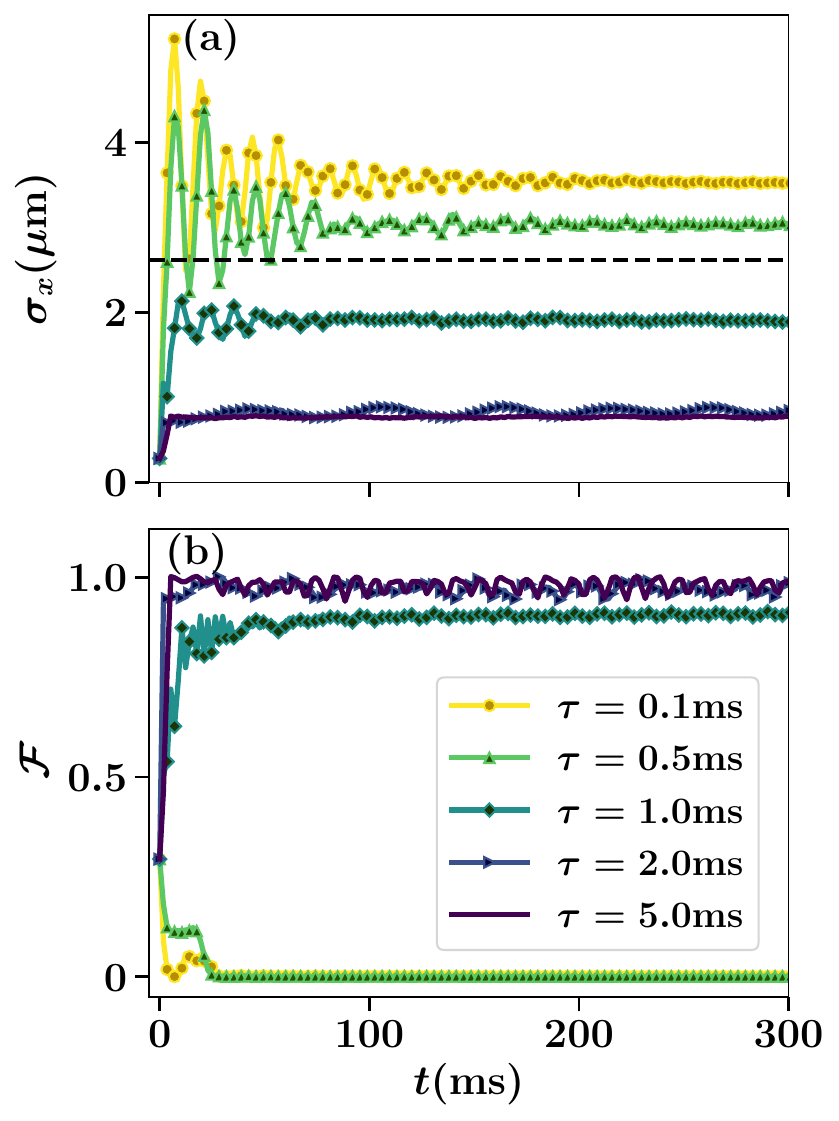}
		\caption{(a) Temporal evolution of the rms width $\sigma_x$ for the SD state subjected to a linear change in the polarization angle with $\Delta \theta = 15^{\circ}$ for different quench times. The black dashed line represents the axial width $\sigma_z$ of the SD ground state. (b) The corresponding fidelity dynamics for the same set of linear ramps, highlighting the effect of the quench time on the elastic to plastic phase transition. The dynamics are shown for five distinct linear ramps associated with different quench times, as indicated in the legend of panel (b).}\label{fig:qt_SD}
	\end{figure}
	
	
	\subsection{Influence of finite quench time on the elastic to plastic phase transition}\label{seciiic}
	So far, we have investigated the effect of sudden changes in the polarization direction on the different phases of a dipolar BEC. We find that the droplet and the SS phases demonstrate an elastic to plastic phase transition when subjected to a sudden change in the polarization direction beyond a critical value. However, in experiments, changing a control parameter such as the polarization direction of the magnetic field may take a small finite time. Therefore, the natural question that arises is: does a similar elastic to plastic phase transition occur under finite-time linear quenches? To investigate this, we consider the SD $(a_s=70a_0)$ and SS $(a_S=90a_0)$ states and perform a linear ramp over different quench time $\tau$ to a final polarization angle to $\Delta \theta = 15^{\circ}$ and $\Delta \theta = 20^{\circ}$, respectively. Note that for an instantaneous quench to the chosen values of $\Delta \theta$, both the SD and SS undergo permanent deformation. For the SD case, we examine the dynamics following the ramp of $\Delta \theta=15^{\circ}$ by plotting the temporal evolution of the rms width $\sigma_x$ and the fidelity for different ramp times $\tau$ in Fig.~\ref{fig:qt_SD}. We find that instead of a sudden quench, even for a small finite quench time up to $\tau_c^{\rm SD} \sim 0.5~\rm{ms}$, the width along the $x$ direction increases above the axial rms width (dashed black line) and the fidelity drops to zero, demonstrating an elastic to plastic phase transition. However, for quench time $\tau>\tau_c^{\rm SD}$, $\sigma_x$ remains smaller than the axial rms width $\sigma_z$ of the SD ground state and, consequently, the fidelity remains close to $1$. This implies that as we slowly change the polarization direction, there is no elastic to plastic phase transition for a large linear ramp time as the atoms of the dipolar BEC adiabatically follow the instantaneous ground state with the modified polarization direction. This is a direct indication that the transition we have indicated requires genuine non-equilibrium dynamics. \par
	\begin{figure}[tb!]
		\centering
		\includegraphics[width=0.46\textwidth]{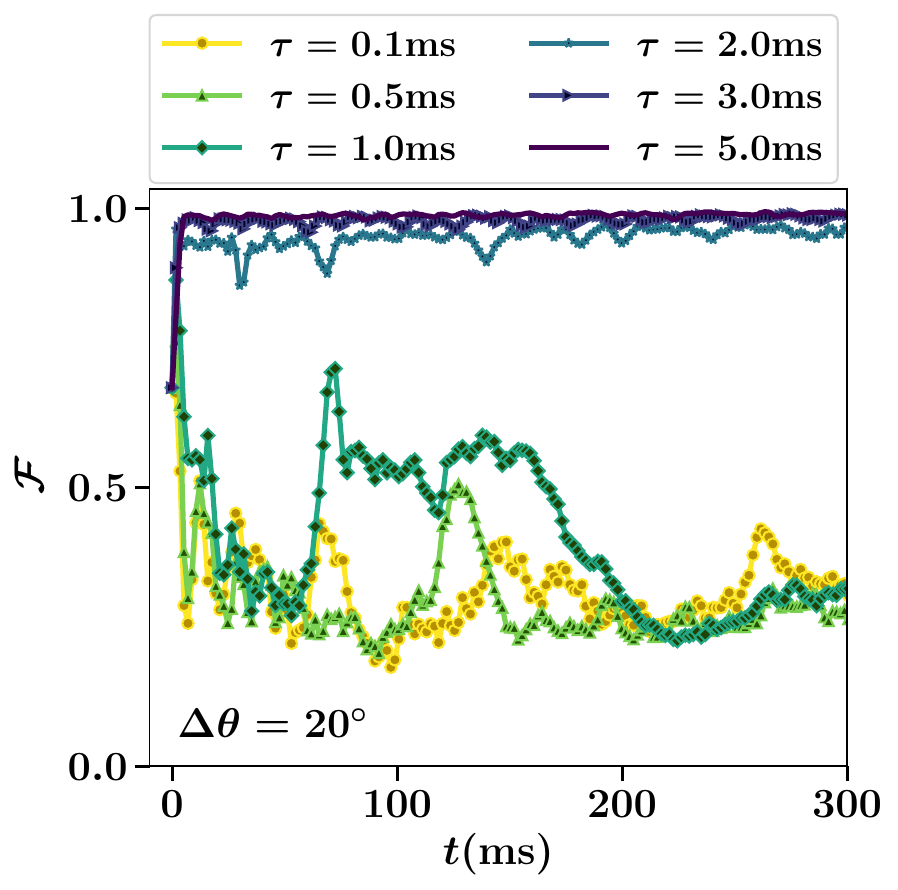}
		\caption{Temporal evolution of the fidelity $\mathcal{F}$ for the SS state following a linear change in the polarization angle with $\Delta \theta = 20^{\circ}$. The dynamics are shown for six distinct linear ramps corresponding to different quench times, as indicated in the legend.}\label{fig:qt_SS}
	\end{figure}
	For the SS case, we examine the dynamics induced by a linear ramp of the external polarizing magnetic field $\Delta\theta=20^{\circ}$ by analyzing the fidelity for six different ramp durations, as shown in Fig. \ref{fig:qt_SS}. For ramp times $\tau \leq \{\tau_c^{\rm SS} \sim 1~\rm{ms}\}$, the fidelity drops below $\mathcal{F} < 0.5$, indicating a change in the crystalline structure, while for quenches with $\tau > \tau_c^{\rm SS}$, the fidelity remains close to unity. The overall behavior closely resembles that of the SD case, with the key distinction that the SS state exhibits a permanent structural deformation for comparatively longer critical ramp times, i.e., $\tau_c^{\rm SS}>\tau_c^{\rm SD}$.\par
	
	The critical ramp time beyond which the fidelity remains close to unity, signifying the absence of an elastic to plastic phase transition, is governed by the typical response time-scale of the system to external perturbations. A heuristic estimate of this time scale can be taken as the inverse of the scissors-mode frequency associated with small angular perturbations. As discussed earlier in Section \ref{seciiia}, the scissors-mode frequency of the SD state with $a_s=70a_0$ is $\Omega_{\rm sc} = 2\pi \times 593 ~\rm{Hz}$, corresponding to an internal time scale of $T_i^{\mathrm{SD}} \approx 0.3$ ms which is close to the numerically observed $\tau_c^{\rm SD}$. In comparison, the SS state exhibits scissors-mode excitations with two different frequencies: one associated with the SF component ($\Omega_{\rm sc}=2\pi\times 130.4~\rm{Hz}$) and the other with the droplet array ($\Omega_{\rm sc}=2\pi\times276.03~\rm{Hz}$). The lowest-frequency mode among these defines the longest internal time scale, $T_i^{\mathrm{SS}} \approx  1.2~\rm{ms}$ which is also approximately equal to $\tau_c^{\rm SS}$. Therefore, when the polarization direction is varied on a time scale much longer than the intrinsic time scale of the SD and SS states $\tau\gg T_i^{\rm{SD/SS}}$, no elastic to plastic phase transition occurs. Thus, the elastic to plastic phase transition arises from an inherently non-adiabatic response of the dipolar BEC to the polarization direction change. In contrast, the SF state displays only excitations of various collective modes under very rapid quenches ($\tau < 1.1~\rm{ms}$), without undergoing any solid-like structural transition, even for large angular changes in the polarization direction.
	
	
	\section{Conclusions}\label{seciv}
	In conclusion, we have investigated the dynamic response of different phases of dipolar BECs, viz. SF, SS and SD states, to the mechanical shear stress induced by applying a sudden change in the polarization direction of the atomic dipoles. For small angular deviation, all the phases of dipolar BEC exhibit a scissors mode oscillation. The scissors-mode frequency in the SF phase is primarily governed by the trapping geometry, remaining nearly unaffected by interparticle interactions or deviation in the polarization angle. This independence signifies a non-elastic response, while the persistence of undamped oscillations throughout the time evolution highlights the inherently non-rigid nature of the SF phase. \par 
	
	In contrast, the droplet and SS phases exhibit interaction-dependent scissors mode frequencies and damping of the oscillation amplitude over time, reflecting an elastic response to the shear stress. The stronger damping and broader spectral width observed in isolated droplets (SD and MD) compared to the SS, indicate a higher degree of mechanical rigidity in isolated droplets. These results demonstrate that the specific excitation scheme employed in this work for the excitation of the scissors-mode can serve as an effective diagnostic tool for probing the rigidity and elastic properties across different quantum phases of dipolar BECs.\par
	
	As we increase the angular deviation, in addition to the scissors-mode, other low-lying modes are excited in the condensate. Although the fidelity of the SF phase oscillates, it remains close to 1, irrespective of the amount of angular deviation. In sharp contrast, the SD state exhibits a pronounced growth of the transverse correlation length under increasing mechanical stress, which eventually exceeds the axial rms width. This signals an instability that drives fragmentation and the emergence of a crystalline lattice structure with multiple droplets. In the SS phase, all droplets undergo similar effects, and beyond a critical angular deviation both the crystalline arrangement and the droplet number are altered, thereby marking an elastic to plastic phase transition. Interestingly, the background superfluid softens the effects of external shearing stress on the SS state, allowing for a greater critical angular deviation compared to the more rigid SD state for the elastic to plastic transition. Finally, we note that supersolid and droplet phases of dipolar BECs with lifetimes of the order of several hundred milliseconds to seconds have been observed in experiments \cite{ferrier-barbut_Observation_2016, poli_synchronization_2025}. Thus, both the characteristic broadened spectral features in the Fourier spectrum of the scissors-mode signal (with damping of the order of few hundred milliseconds for the droplet phase) as well as irreversible structural rearrangement for large angular deviations should be experimentally observable.\par
	
	Our work provides a new pathway to explore the mechanical response of different phases of dipolar BECs and opens a plethora of future research directions. One straightforward direction would be to explore the effect of finite temperature \cite{sanchez-baena_Heating_2023, he_Accessing_2025} on the shear rigidity of dipolar SS and isolated droplet states. Finite-temperature effects may also drive lattice melting \cite{bombin_Creating_2025} and defect formation, providing deeper insight into the stability and robustness of solid-like order in these systems. Finally, the sensitivity of the state of the dipolar BEC to the polarization direction can potentially be exploited to design vector magnetic field sensors \cite{sensor1,sensor2}.
	
	\section*{Acknowledgements}
	S.H. and B.P.V. acknowledge funding from the DST National Quantum Mission through the project
	DST/FFT/NQM/QSM/2024/3. B.P.V acknowledges support from MATRICS Grant No. MTR/2023/000900 from Anusandhan National Research Foundation, Government
	of India. H.S.G. gratefully acknowledge the support from the Prime Minister's Research Fellowship (PMRF), India. A.P. acknowledges financial support by the Deutsche Forschungsgemeinschaft (DFG, German Research Foundation) via the Collaborative Research Center SFB/TR185 (Project No. 277625399).
	\appendix
	
	
	\section{Linear response theory and Sum rule}\label{A}
	At $T=0$K, the dynamic structure factor relative to the response of a certain operator $\hat{F}$ is given by
	\begin{equation}
		S(\hat{F},\omega)=\sum_e\abs{\mel{e}{\hat{F}}{0}}^2\delta(\hbar\omega-\hbar\omega_e) \label{S_F},
	\end{equation}
	where $\ket{0}$ corresponds to the ground state, $\ket{e}$ represents the excited eigenstates and $\omega_e$ is the transition frequency from the ground state to the excited state $\ket{e}$. Although it is difficult to evaluate $S(\hat{F},\omega)$ starting from the many-body Hamiltonian of our considered system, useful insights can be extracted from the energy-weighted moments of the dynamic structure factor. The generalized form of the $n$th-order energy-weighted moment can be written as
	\begin{align}
		m_n(\hat{F})=\hbar^{n+1}\int \dd{\omega} \omega^n S(\hat{F},\omega)\label{m_n}.
	\end{align}
	By substituting Eq.~\eqref{S_F} in Eq.~\eqref{m_n}, we obtain
	\begin{align}
		m_n(\hat{F})=\hbar^{n+1}\sum_e \omega_e^n \abs{\mel{e}{\hat{F}}{0}}^2\label{m_nF}.
	\end{align}
	In our case, we perturb the system by a sudden change in the polarization direction, which generates the $y$-component of the angular momentum ($\hat{L}_y$) and leads to excitations in the system. When the amplitude of deviation of polarization direction is small, then we can apply the linear response theory and from Eq.~\eqref{m_nF} with $\hat{F} = \hat{L}_y$ we obtain
	\begin{align}
		\frac{m_1(\hat{L}_y)}{m_{-1}(\hat{L}_y)}=\frac{\hbar^2\sum_e \abs{\mel{e}{\hat{F}}{0}}^2\omega_e }{\sum_e\abs{\mel{e}{\hat{F}}{0}}^2\omega_e^{-1}}\label{m1/m-1}.
	\end{align}
	Under the application of the $\hat{L}_y$ operator, the scissors mode is the lowest excited mode with frequency $\omega_e^{\rm {min}}=\omega_{sc}$. From Eq.~\eqref{m1/m-1}, one can obtain the rigorous upper bound of the scissors mode oscillation
	\begin{equation}
		\hbar\omega_{sc}\leq\sqrt{\frac{m_1(\hat{L}_y)}{m_{-1}(\hat{L}_y)}},
	\end{equation}
	with the equality holding when the $\hat{L}_y$ operator only excites the lowest mode. Now the energy-weighted moment $m_1(\hat{L}_y)$ can be expressed as 
	\begin{equation}
		m_1 (\hat{L}_y)= \frac{1}{2} \langle [\hat{L}_y, [\hat{H}, \hat{L}_y]] \rangle,\label{Am_1}
	\end{equation}
	where $\hat{H}$ is the atomic many-body Hamiltonian. All the terms of the Hamiltonian commute with the $\hat{L}_y$ operator, except the DDI  and the trapping potential terms. Therefore Eq.~\eqref{Am_1} becomes,
	\begin{align}  
		m_1(\hat{L}_y) = \frac{1}{2} \left(\langle [\hat{L}_y, [\hat{V}_{\rm dd}(\vb{r}-\vb{r^{\prime}}), \hat{L}_y]] \rangle + \langle [\hat{L}_y, [\hat{V}_{t}(\vb{r}), \hat{L}_y]] \rangle\right),\label{eq:m1Ly}
	\end{align}
	where $\hat{L}_y=\sum_i z_ip_x^i-x_ip_z^i$ is the angular momentum operator, the 3D harmonic trapping potential, experienced by all the atoms in the condensate is given by
	\begin{equation}
		\hat{V}_t (\vb{r}) = \sum_{i=1}^{N} \frac{1}{2}m\big[\omega_{\perp}^2(x_i^2+y_i^2)+\omega_z^2z_i^2)]\big],
	\end{equation}
	with  $\omega_{\perp}=\omega_x=\omega_y$ $(\omega_z)$ denoting the transverse (axial) trapping frequencies, and the dipole-dipole interaction (DDI) potential is 
	\begin{equation}
		\hat{V}_{\rm{dd}}=\frac{1}{2} \sum_{i,j} \frac{\mu_0\mu_m^2}{4\pi \abs{\vb{r}_i-\vb{r}_j'}^3}\Big[1-3\frac{(z_i-z_j)^2}{\abs{\vb{r}_i-\vb{r}_j}^2}\Big].
	\end{equation}
	Note that in the above expression, we have taken the initial configuration with all the dipoles polarized along the $z$-axis (set by the external magnetic field).
	\begin{figure}[tb!]
		\centering
		\includegraphics[width=0.48\textwidth]{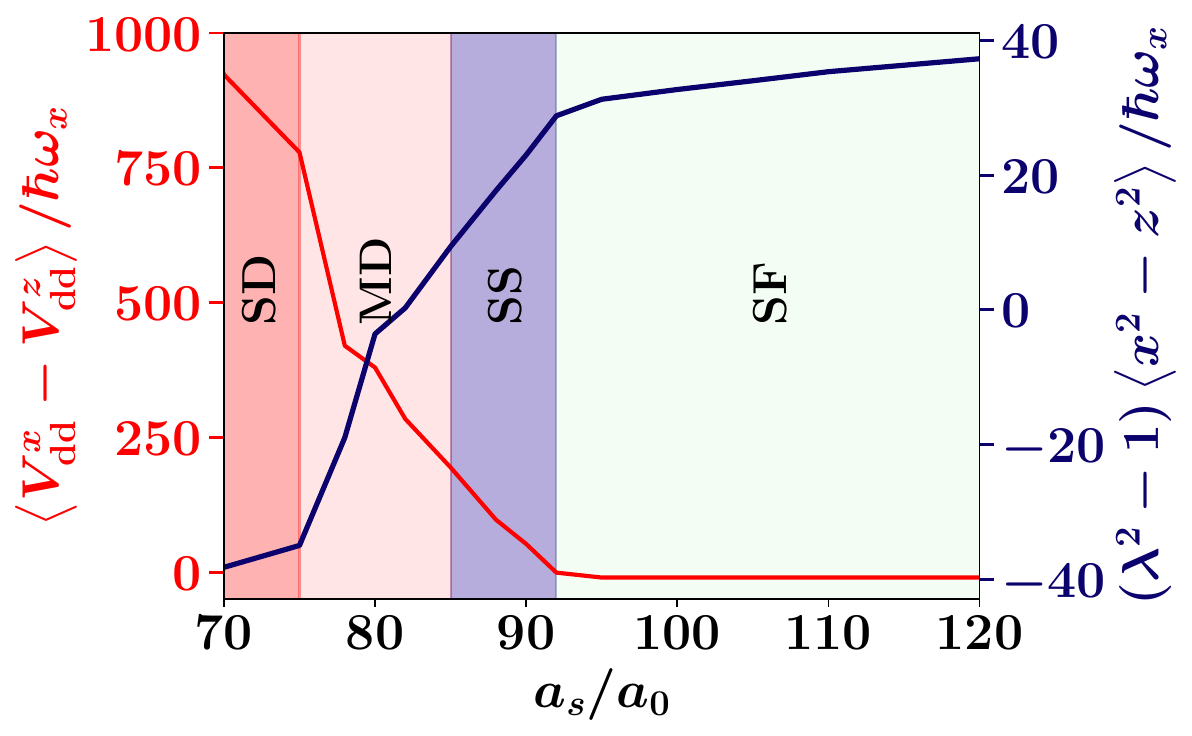}
		\caption{Variation of the restoring force of the scissors mode oscillation arising from dipolar anisotropy $\expval{V_{\rm dd}^x-V_{\rm dd}^z}$ and trapping anisotropy $(\lambda^2-1)\expval{x^2-z^2}$ as a function of the $s$-wave scattering length, where $\lambda=\omega_z/\omega_{\perp}>1$.}\label{fig:A1}
	\end{figure}
	The commutator involving the DDI potential and the $y$-component of angular momentum in Eq.~\eqref{eq:m1Ly} gives
	\begin{equation}
		[\hat{V}_{\rm dd}, \hat{L}_y]=i\hbar\sum_{i,j}\frac{3\mu_0\mu_m^2}{4\pi \abs{\vb{r}_i-\vb{r}_j}^5}(x_i-x_j)(z_i-z_j).
	\end{equation}
	Similarly, the second commutator gives
	\begin{equation}
		[\hat{L}_y, [\hat{V}_{\rm dd}, \hat{L}_y]]= \hbar^2 \sum_{i,j} \frac{3\mu_0\mu_m^2}{4\pi \abs{\vb{r}_i-\vb{r}_j}^5}\{(z_i-z_j)^2-(x_i-x_j)^2\}.
	\end{equation}
	Therefore, in the mean-field framework, the first term in Eq. \eqref{eq:m1Ly} can be written as
	\begin{equation}
		\langle[\hat{L}_y, [\hat{V}_{\rm dd}, \hat{L}_y]] \rangle=\hbar^2\langle V_{\rm{dd}}^x -V_{\rm{dd}}^z\rangle,\label{eq:c1}
	\end{equation}
	where
	\begin{equation}
		\langle V_{\rm dd}^\eta \rangle = \int \dd{\vb{r}} \dd{\vb{r}}' n(\vb{r}) V_{\rm dd}^\eta(\vb{r} - \vb{r}') n(\vb{r}'), \, \eta\in\{x,y,z\}
	\end{equation}
	and
	\begin{equation}
		V_{\rm{dd}}^\eta(\vb{r} - \vb{r}')= \frac{\mu_0\mu_m^2}{4\pi \abs{\vb{r}-\vb{r'}}^3}\left[1-3\frac{(\eta-\eta')^2}{\abs{\vb{r}-\vb{r'}}^2}\right].
	\end{equation}	
	The commutator involving the trapping potential results
	\begin{align}
		[\hat{V}_t,\hat{L}_y]=i m\hbar (\omega_x^2-\omega_z^2)\sum_i x_iz_i.
	\end{align}
	This implies that in an anisotropic harmonic trap, angular momentum is coupled to the quadrupole degrees of freedom. Similarly, the second commutator involving the trapping potential leads to the following expression
	\begin{align}
		\langle[\hat{L}_y,[\hat{V}_t,\hat{L}_y]\rangle=m\hbar^2 (\omega_z^2-\omega_x^2)\langle x^2-z^2\rangle\label{eq:c2}.
	\end{align}
	Putting Eq.~\eqref{eq:c1} and \eqref{eq:c2} in Eq.~\eqref{eq:m1Ly}, we obtain
	
	\begin{equation}
		m_1(\hat{L}_y)=\frac{\hbar^2}{2}\left[\langle V_{\rm{dd}}^x-V_{\rm{dd}}^z\rangle+m(\omega_z^2-\omega_x^2)\langle x^2-z^2\rangle\right]\label{eq.m_1}.
	\end{equation}

	\begin{figure*}[htb!]
		\centering
		\includegraphics[width=\textwidth]{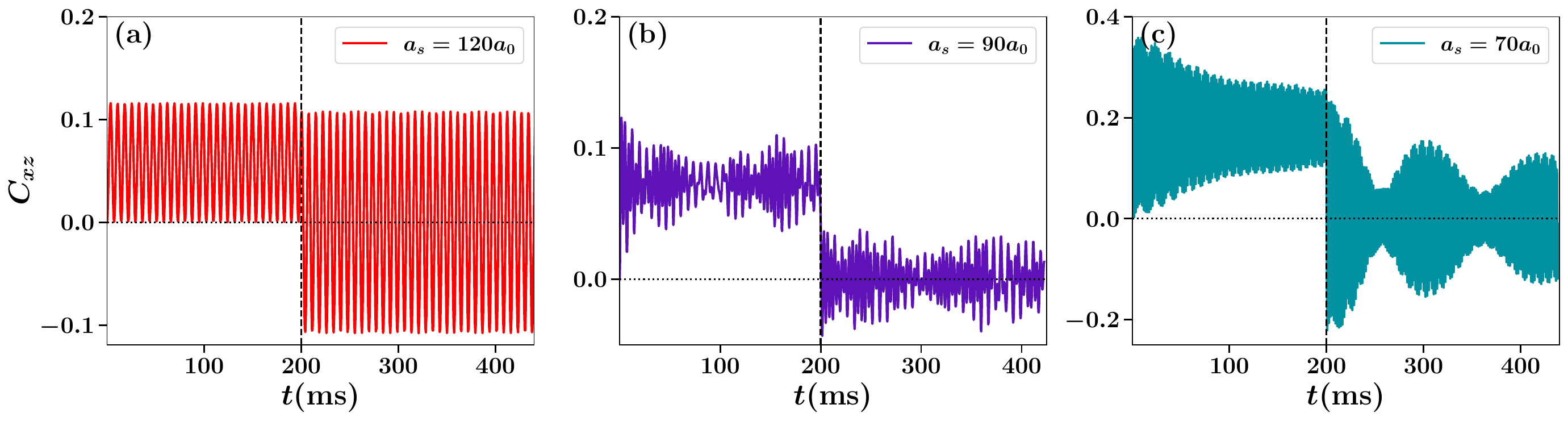}
		\caption{The panels (a-c) demonstrate the temporal evolution of the scissors mode in the $x$-$z$ plane for (a) the SF ($a_s=120a_0$), (b) SS ($a_s=90a_0$) and (c) the SD ($a_s=70a_0$) state, respectively, following a cyclic polarization reorientation by $\Delta \theta=2^{\circ}$. The black dashed vertical line indicates the time at which the field orientation is restored to its initial alignment.}\label{fig:8}
	\end{figure*}

	The energy-weighted moment $m_1$ acts as an effective restoring force for the scissors-mode oscillation. As evident from Eq.~\eqref{eq.m_1}, in our case $m_1$ is determined not only by the trapping geometry but also by the anisotropy of the DDI between the atoms in the condensate.
	Fig. \ref{fig:A1} illustrates how the two contributions to $m_1$ arising from the DDI and the anisotropic trapping potential vary with the $s$-wave scattering length. In the SF phase, for large $s$-wave scattering lengths, the restoring force is predominantly governed by the trapping potential with only a minor contribution arising from the anisotropy of the DDI. As the $s$-wave scattering length decreases and the system enters the SS regime, the influence of the dipolar interaction becomes increasingly significant in determining the restoring force. Upon further reduction of $a_s$, the dipolar anisotropy $\expval{V_{\rm dd}^x-V_{\rm dd}^z}$ eventually dominates over the effect of anisotropy in the trapping potential $(\lambda^2-1)\expval{x^2-z^2}$ in determining the restoring force of the scissors mode oscillation in the droplet (SD/MD) phases.


	\begin{figure*}[htb!]
		\centering
		\includegraphics[width=0.94\textwidth]{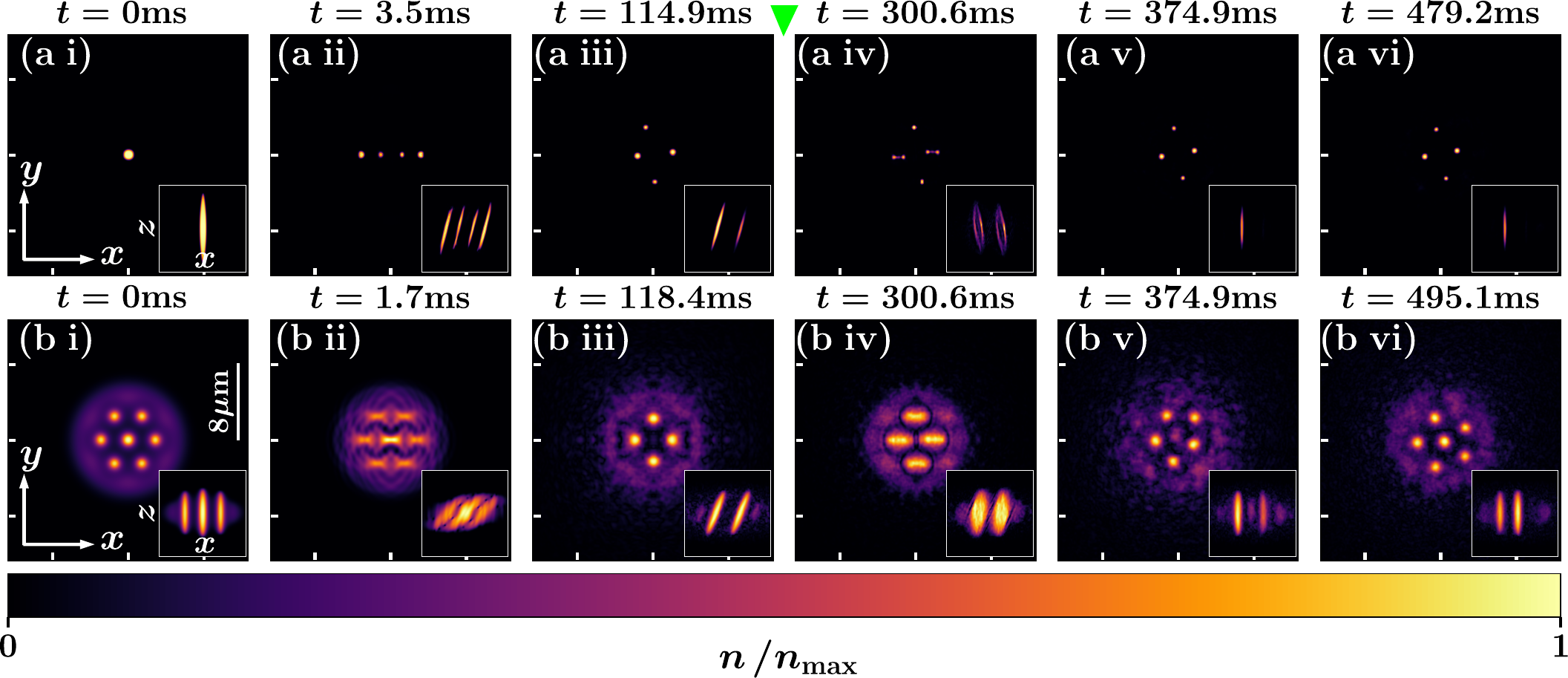}
		\caption{Panels (a i–a vi) and (b i–b vi) show snapshots of the density profiles in the $z=0$ plane for (a) the SD ($a_s=70a_0$) and (b) the SS ($a_s=90a_0$) state, respectively, following a double change in polarization direction of the external magnetic field by a large angle $\Delta \theta>\Delta \theta_c$. The insets show the corresponding density distribution in the $y=0$ plane. The green triangular marker indicates the time ($t=300 ~\rm{ms}$) after which the field orientation is restored to its initial alignment. The results correspond to $\Delta \theta=15^{\circ}$ for the SD state and $\Delta \theta=20^{\circ}$ for the SS state.}\label{fig:9}
	\end{figure*}

	\section{Double orientation quench}\label{B}
	In the main text, we discussed the effects of small and large shear stresses generated by instantaneous and linear changes in the orientation of the magnetic field on different phases of a dipolar BEC. In this section, we consider a scenario where the system is initially prepared in various phases of a dipolar BEC with the magnetic field polarized along the $z$-axis. The field orientation is first tilted by an angle $\Delta \theta$, and after a certain duration, the polarization direction is restored to its initial alignment.\par 

	In Fig.~\ref{fig:8}, we show the time evolution of $C_{xz}(t)$ for the SF, SS, and SD states following a double change in orientation with a small $\Delta \theta=2^{\circ}$. For such a small angular deviation $\Delta \theta$, all three phases initially exhibit scissors-mode oscillations about $C_{xz}\neq0$, consistent with the behavior discussed in Sec.~\ref{seciiia}, arising from the shear stress generated by the sudden change in orientation. Once the field is restored to its initial direction at $t=200~\rm{ms}$, all phases oscillate around $C_{xz}=0$. The SF state shows undamped oscillations, with the amplitude of the scissors mode even increasing after the second orientation quench and being sustained over time [see Fig. \ref{fig:8}(a)]. Consequently, the SF phase does not recover its initial static configuration even after the polarization is realigned, indicating the absence of rigidity. In contrast, the SS and SD states exhibit damped oscillations that gradually relax toward the initial polarization direction, demonstrating an elastic response within the elastic regime [see Figs. \ref{fig:8}(b) and \ref{fig:8}(c)]. The amplitude of the scissors-mode oscillation decays faster in the droplet state than in the SS state, indicating that the SS phase requires a longer time to return to its initial static configuration. This behavior suggests that the droplet phase possesses greater rigidity compared to the SS state.\par
	
	For large angular deviations, $\Delta\theta > \Delta\theta_c$, both the droplet and SS states undergo permanent deformation. Here, we demonstrate a scenario in which the SD and SS states are subjected to sudden angular deviations of $\Delta\theta=15^{\circ}$ and $\Delta \theta=20^{\circ}$, respectively. Under this sudden change in polarization direction, the SD state fragments into a MD configuration, while the SS state experiences a change in its crystalline structure and number of droplets [see Figs. \ref{fig:9}(a i- a iii) and \ref{fig:9}(b i- b iii)]. At $t=300~\rm{ms}$, once the magnetic-field polarization is realigned along the axial direction, i.e., the initial orientation, the droplets in SD and SS states subsequently orient along the field, but the deformation persists. After the field is restored, the SD state, having transitioned into the MD configuration, remains in the deformed MD configuration, with droplets oriented along the direction of the external field [see Figs. \ref{fig:9}(a iv- a vi)]. Similarly, in the SS state, the crystalline lattice structure altered by the induced shear stress does not revert to its initial configuration [Figs. \ref{fig:9}(b i- b iV)]. The shearing stress produced during the sudden realignment process of the external magnetic field leads to a change in the number of droplets and crystalline structure for a sufficiently large $\Delta\theta\geq20^{\circ}$ [see Figs. \ref{fig:9} (b iv- b vi)]. However, the SS state does not return to its ground state configuration with a polarizing magnetic field along the axial direction [Fig. \ref{fig:9} (b i)], instead exhibiting a permanent structural deformation [see Fig. \ref{fig:9} (b vi)].

	\bibliographystyle{apsrev4-2}
	\bibliography{reference} 
	
\end{document}